# Manufacturing and Assembly for Vacuum Technology

*S. Atieh, G. Favre, S. Mathot*
CERN, Geneva, Switzerland

**Abstract**
In order to satisfy ultrahigh-vacuum requirements several mechanical engineering disciplines need to be pushed to their limits. At the manufacturing level high precision turning or diamond turning is necessary to obtain smooth surfaces, which are further improved by special machining. During assembly, in particular during welding and brazing one must take care not to damage the surface quality. Several assembly techniques are shown and illustrated with examples.

## 1   Introduction

This paper describes commonly used techniques for machining and assembly of high vacuum components. The purpose is to give general information on these techniques and to present the advantages and limitations when applied to the field of high vacuum systems. The aim is not to propose a complete catalogue of all possible techniques but rather to focus on the more efficient and proven ones, disregarding exotic and rarely used methods as well as not applicable or not recommended processes for high vacuum applications.

Present trends and future perspectives are also discussed and the paper concludes with some recommendations based on the experience gained at CERN on the construction of vacuum chambers and accelerator components.

## 2   Manufacturing

Already in 1983, Taniguchi presented a curve depicting the general improvement of machining accuracy with time for a period covering much of the twentieth century [1]. This curve fits well the situation in 2017, e.g., nanometre accuracy for ultra-precision techniques such as ion beam machining or soft X-ray lithography.

These techniques are outside of the topic of this paper, in which we will discuss only high precision methods such as turning and milling, where the ultimate precision is in the range of a tenth of a micron. However, we prefer to discuss the shape accuracy. We will see that we are then in the range of tens of micrometres for high precision techniques and in the micron range using diamond tools.

For RF [Radio-Frequency] cavities, in addition to the shape accuracy, the surface roughness is also an important parameter to take into account. An experimental rule determines that the surface roughness must be a quarter of the skin depth, which varies as the square roots of the inverse of the RF frequency and material conductivity. Then, considering copper surfaces, for an RFQ [Radio Frequency Quadrupole] working at 350 MHz the surface roughness (Ra) must be 0.8 micrometres, 0.4 micrometres for an accelerating structure working at 3 GHz and less than 100 nanometres for high frequency X-band electron cavities (8–12 GHz) [2].

However, the typical surface finishes for classical machining operations have been slightly improved during the last decades. In the 1990s, the better Ra values are 0.025 micrometres for the diamond turning method, and this value is still valid. Notable improvement is for diamond milling where we can obtain now a similar finish quality.

## 2.1 High precision turning

This method is used for axisymmetric components. For machining, the parts are rotated and a cutting tool can remove material by moving along the axis of rotation and perpendicular to this axis. It is important to note that this is a continuous process where a single edge removes all of the material.

A careful control of the chip is important for the reliability and the work piece quality. The main limitation comes for the so-called 'chatter' effect, which correspond to a self-excited vibration caused by the interaction of the machine tool and the chip removal process. These vibrations limit the surface finish and accuracy.

The achievable performance on state-of-the-art equipment is a shape accuracy in the range of 10 micrometres and a surface roughness (Ra) less than 0.2 micrometres (on OFE [Oxygen Free Electronic] copper).

## 2.2 Diamond turning and milling

An improved precision can be obtained when a diamond replaces the standard hard metal cutting tool. This is possible not only for turning, but also for milling (see below). It should be noted that it is only at the tip of the cutting tool that a single or a poly-crystal diamond having a high precision shape is fixed. Also, for high precision, changing the tools is not sufficient. The turning or milling machine must have the necessary stiffness and precision and are generally dedicated for this process.

With the diamond turning or milling technique, one can achieve a shape accuracy in the micron range and a surface finishing (Ra) of less than 100 nanometres (Fig. 1).

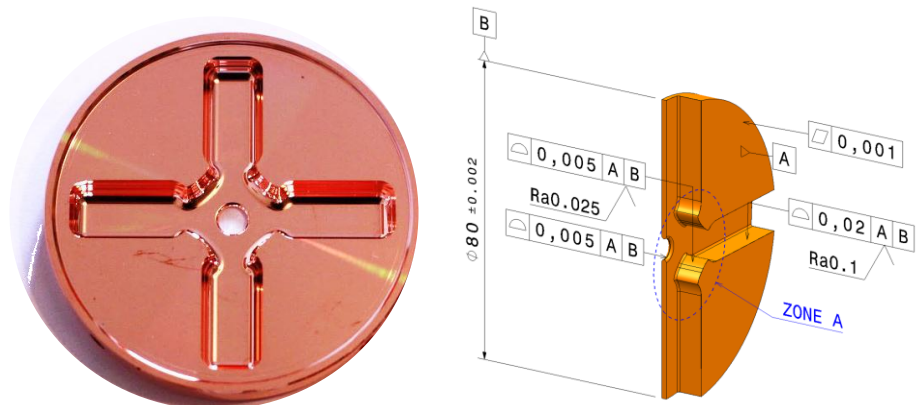

**Fig. 1:** Example of ultra-precision diamond machined piece

An important point to note is that diamond cutting is not possible for ferrous material machining, including stainless steel. This is due to a chemical reaction between diamond and iron that causes a catastrophic wear of the tool that is unstable and impossible to predict.

## 2.3 High precision milling

This machining method allows for the fabrication of prismatic shapes. The piece is generally fixed and a rotating tool moves in contact with the piece to remove the material.

The process is an interrupted cutting operation; one or several edges cut the material alternatively producing many impacts. Thermal issues and mechanical stress can be important. Five-axis machining is possible for complex shapes.

Achievable performance with state-of-the-art milling machines corresponds to a shape accuracy between 10 and 20 micrometres and a surface roughness of about 0.8 micrometres for stainless steel and 0.2 micrometres for copper.

In milling, and also in turning, the affected layer during the machining must be considered, especially if coating is foreseen for the final piece or if a high electric field is present. In the affected layer, between a few micrometres to about 100 micrometres, the material mechanical properties are modified. The depth affected by these modifications depends strongly on the machining conditions and the material properties. Studies are ongoing at CERN to characterize the affected layer and to improve the machining parameters. High precision electron microscopy techniques and nano hardness measurements are used to measure the density of dislocations that are more significant under the surface than in the bulk.

## 2.4 Machine centre

New machines, called 'machine centres' are a new class of machines made for multitasking work and are capable of turning and milling on a same piece. A variety of possible movements for the piece and for different tools allows a variety of shapes to be achieved in a single machining step.

Two main orientations are used. Vertical machine centres are generally used for precision and complex machining. Horizontal machine centres are generally used more for large productions with not only tools but also a piece automatic changer.

## 2.5 Ceramic machining

Ceramics are an important class of materials used for vacuum applications and for accelerator parts. The most commonly used is alumina, a white ceramic made of pure alumina grains ($Al_2O_3$) with a glass binder made of several oxides. This material is an ideal choice for thermal insulations, electrical feedthroughs or RF windows. Silicon nitride (SiC) is used as an RF absorber, aluminium nitride (AlN) allows the exceptional combination of a good electrical resistivity with a good thermal conductivity; pure alumina (Sapphire) and diamond are used for optical or RF windows.

All these ceramics are very hard and difficult to machine. Special tools made of diamond powder sintered on copper are used for milling and turning. While for milling the tools are of similar shapes to those used for metal works, on a lathe machine an additional system is used to rotate the cutting tool.

For some applications, machinable ceramics can be used. These materials are made of glass and ceramic (glass-ceramic or vitroceramic) and can be machined with conventional (hard metal) cutting tools. Commercial names are Macor, an alumina based, and Shapal, an AlN based, glass-ceramic. For vacuum applications, a heat treatment of the Macor is recommended.

## 2.6 Cutting fluids

Machining of components for vacuum applications and accelerators needs constant precautions. A parameter often forgotten is the cutting fluid. Some cutting fluids can be extremely difficult to clean (remove) or could induce long-term corrosion, compromising the use of the component manufactured or affecting a complete system.

A general recommendation is to select an oil-free or water soluble cutting oil without additives containing sulphur, chlorine, zinc, or phosphor. The best practice is to qualify the surface treatment to be used by outgassing and surface analysis. If possible, a good option is to perform the final machining (finishing) without fluid or using pure ethanol as the cutting fluid (Fig. 2).

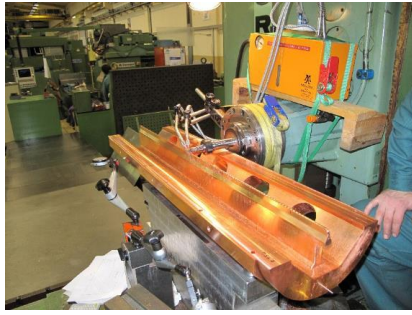

**Fig. 2:** Example of finishing with pure ethanol lubrication

## 2.7 Avoided or less appropriate techniques

Some manufacturing processes must be considered with caution for vacuum applications. These include polishing, which should take place only with appropriate materials as SiC, alumina, or diamond (in the case of an RF field, only diamond can be accepted). Water, laser, or plasma cutting are only for rough machining.

Electrical discharge machining (EDM) can be problematic if the cutting wire contains zinc (brass) or because the surface state after machining is not ideal for vacuum applications.

## 2.8 Sheet metal forming

Sheet metal forming is an important fabrication process for vacuum and accelerator components. Spinning, pressing, deep drawing, or hydroforming are commonly used to obtain complex vacuum chamber or accelerating cavities.

Good examples are multi-cell copper or niobium cavities produced from a single seamless tube by hydroforming. This method can be used also for the fabrication of aluminium bellows. Larger cavities are manufactured by spinning and electron beam welding: half-cells are formed from thin flat metal discs pushed onto a rotating former and welded side by side. Extrusions, on a tube or a metal sheet, can significantly reduce the deformations and can make welding operations easier.

# 3 Assembly

## 3.1 TIG welding

TIG (tungsten inert gas) welding or, more correctly, GTAW (gas tungsten arc welding) is the most commonly used assembling technique for metals and mainly for stainless steel, an alloy often used for vacuum vessels. In front of a joint between two metals parts, a tungsten electrode induces an electrical arc that melt the two metals and, after cooling, produces the weld. The electrode is not eroded and an inert gas flow (generally argon) protects the metal against oxidation. Depending on the configuration, a rod of filler metal can be used to bring material into the melt. The movement of the electrode along the joint produces continuous and vacuum tight welding.

The process can be manual, i.e., the welder maintains and moves the welding head, or automatic with a programmed movement of the head or of the piece and an automatic adjustment of the welding parameters. Experienced welders can achieve using TIG perfectly tight joints without internal defects.

This is possible only for the welding of high quality metals. TIG is effectively sensitive to impurities in the base materials. Good results mean a careful qualification of the metals, the welders, and the process. A strong limitation is also the assembly of dissimilar metals. Only similar metals or a similar grade of alloys can be weld together.

For high vacuum and cryogenic applications, additional aspects are relevant. Austenitic stainless steel grades are generally used and are easily welded. However, austenitic stainless steels may contain a residual amount of delta ferrite, which can reduce the toughness of the weld. Control of the ferrite is required for critical applications. Also, hot cracking in the weld is dependent on the impurity level. If a high quality stainless steel (316LN grade) is mixed with lower quality grades (304L or 316L) the dilution of impurities can be the cause of hot cracking (cf. S. Sgobba presentation).

Welding of vacuum vessels imposes certain rules. Generally speaking, the weld must always be on the vacuum side. Air side welding for mechanical reinforcement can be accepted. However, in that case the welding must be discontinue. These design aspects are extremely important to avoid the formation of pockets and possible virtual leaks. Regions close to a weld that cannot be efficiency rinsed after cleaning are also sources of issues. This is even more critical if more aggressive chemical treatments (e.g., electroplating) are used.

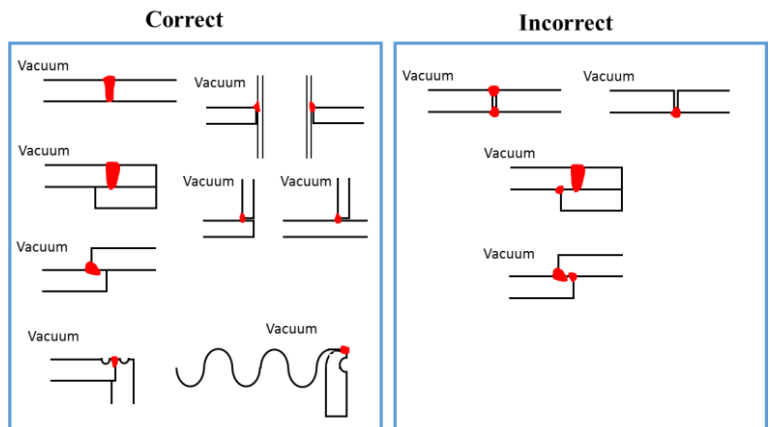

**Fig. 3:** Correct and incorrect welding configurations for vacuum chamber

## 3.2 MIG welding

MIG welding means metal inert gas welding or more correctly GMAW (gas metal arc welding). The main difference compared to the TIG method is that the tungsten electrode is replaced with a consumable metal wire.

This method allows the filling of wider gaps and welding of large joints. MIG is then generally used for large assemblies. The welding quality is (generally) poorer than with TIG.

## 3.3 Electron and laser welding

For these methods, an electron or a laser beam is used to melt the metals. The main difference between the two is that for electron beam welding (EBW) the process is done in vacuum while for laser welding an inert gas protection is used.

The penetration depth is limited for the laser welding and this method is then more adapted for high precision welding of small pieces. Using only a gas protection, this technique is also possible for large components, e.g., the welding of the capillary tubes of the 18 metre long LHC [Large Hadron Collider] beam screens (Fig. 4).

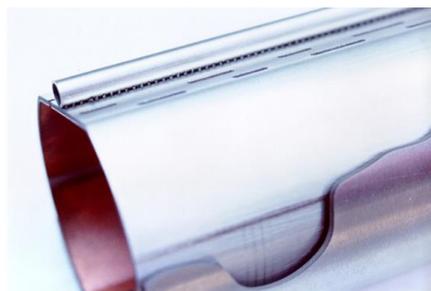

**Fig. 4:** Laser welding for the LHC beam screen, about 500 spots per metre

For electron beam welding, the penetration depth can be very high. Welding of large pieces is limited by the size of the vacuum chamber available for the electron beam welding machine. For components with a simple shape, e.g., a tube, a feedthrough can be used to only partially insert the piece into the vacuum chamber.

EBW allows high precision and high quality welding with a minimum of deformations of the assembly for demanding accelerator equipment such as copper or niobium cavities. Also, the welding process is fast and the heating induced limited, even if the pieces are in vacuum. The process can be then used close to a braze joint and the combination of the vacuum brazing and EBW techniques is commonly used in the field of accelerators.

The previous recommendations regarding the quality of the base metals and the welding of dissimilar materials must be reinforced for the laser and electron beam welding. These methods are extremely sensitive and the qualification of the process and of the material is essential.

### 3.4 Brazing and soldering

Brazing and soldering techniques consist of melting a third material between the two pieces to be assembled. After cooling and solidification of this material, called braze, solder, or filler alloy, we can obtain a solid and vacuum tight joint. By convention, if the process is performed at a temperature below 450°C, we are talking about soldering and solder alloys, with brazing and brazing alloys above this temperature.

The principle is the same for the two techniques, the pieces are heated and the temperature stabilized just below the melting of the solder or braze alloy to homogenize the temperature of the pieces (furnace heating) or of parts of the pieces (direct heating). The homogenization time varies from a few seconds or minutes for the solder of small components, to a few hours for the brazing of heavy pieces. After this the temperature is increased to melt the solder or braze alloy for a period of time that must be as short as possible, a few seconds to a few minutes. After cooling, the pieces are assembled.

Even if the melting point of the joined pieces is not reached during brazing or soldering, the properties of the materials can be affected by the heat treatment and the surface modified by diffusion. A major advantage of these techniques is the possibility to assemble very different metals and also metals and non-metals such as ceramics. A second advantage is the possibility of performing high precision joining with a minimum of deformations.

The joining process is only possible if the filler metal can interact with and wet with efficiency the surface. This is not possible on a piece that is heavily oxidized due to heating in air. For air soldering and brazing, the pieces or the joint must be protected. Generally a mixture of organic or mineral salts, called flux, is used. These have the capability to clean and reduce the oxides at the surface of the pieces. The cleaning of the flux residues can be difficult and, generally, air soldering or brazing is not recommended for high vacuum systems.

However, when the brazing or soldering is performed in a vacuum or reductive atmosphere furnace and when low vapour pressure filler metals are used, the assemblies are compatible with high vacuum applications.

#### 3.4.1 *Vacuum brazing*

In vacuum brazing, not only the pieces are protected against oxidation during the heating but the vacuum treatment helps in the cleaning of the surfaces. Indeed, for copper, we can calculate that the copper oxide ($Cu_2O$) will be unstable at 800°C if the oxygen partial pressure is less than $1.3 \times 10^{-6}$ Torr. This oxygen partial pressure is easily obtained in a vacuum furnace and this explains why we can have a very good brazing on copper in vacuum at 800°C and why the copper surfaces are so bright just after the treatment.

For $Al_2O_3$, the oxygen partial pressure for the equilibrium at 800°C is $2.5 \times 10^{-41}$ Torr. This oxide will be never reduced in a vacuum treatment. The equilibrium values at this temperature for the nickel and the chromium oxides are respectively $1.3 \times 10^{-16}$ and $1.3 \times 10^{-30}$ Torr. These oxygen partial pressures are also never reached in a vacuum furnace but we do obtain very good brazing on nickel platted surfaces and also on stainless steel if the right filler metal is used. The mechanisms allowing in-vacuum brazing and good wetting of a liquid metal for a large number of metals despite the natural oxide layer is not yet fully understood. However, the result is that the quality of the wetting of the filler metals can be high enough to allow for brazing on large surfaces. The wetting is also generally sufficient to obtain good joins on a large number of metals and alloys used for vacuum and accelerator applications.

The drawbacks of this technique are the tight mechanical tolerances needed for the assemblies and, as mentioned, the possible deterioration of the properties of the materials due to the heat treatment at high temperature. For the commonly used silver-based filler alloys, the gap between the pieces must be as low as 25 μm. For other alloys used at higher temperatures, such as gold-copper of nickel-chromium alloys, the gap can be up to 100 μm.

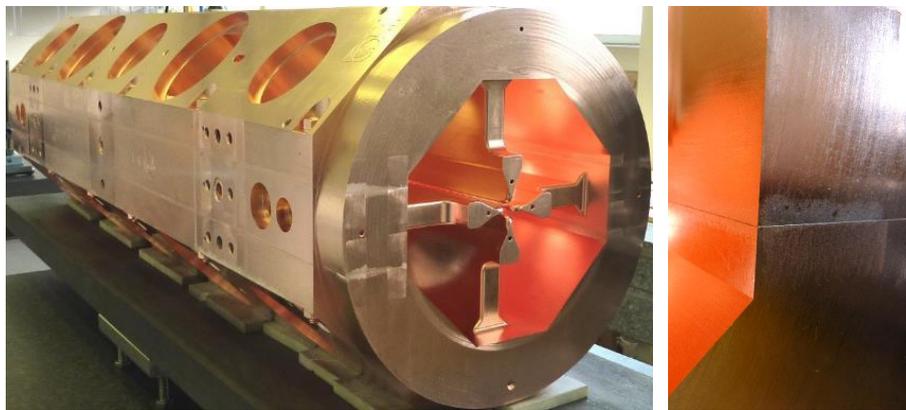

**Fig. 5:** Example of high precision vacuum brazing of copper. This one metre long, 400 kg, RFQ is assembled with a precision of less than 20 μm.

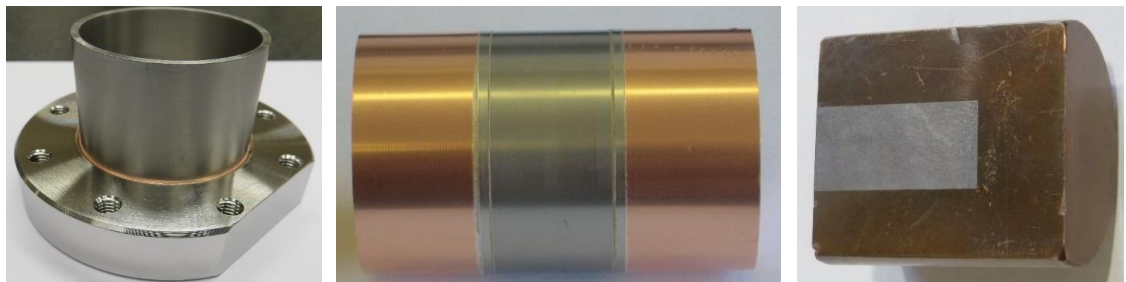

**Fig. 6:** Example of dissimilar metal vacuum brazing: niobium-stainless steel (left), copper-stainless steel-titanium (centre). and copper-tungsten (right).

### 3.4.2    *Vacuum brazing of alumina*

Alumina is a very common ceramic used in vacuum and accelerator applications. Alumina is made of pure $Al_2O_3$ grains sintered with a mixture of other oxides ($SiO_2$, MnO, CaO, etc.). The purity of aluminium oxide is generally between 95 to 99%. An efficient way to braze alumina is to metallize a part of the surface using the moly-manganese process.

This method consists of the application of a mixture of molybdenum powder and oxides on the alumina surface. Then, a heat treatment above 1200°C in a wet hydrogen atmosphere induces the reduction of the molybdenum and the interaction between the oxides from the mixture and from the alumina. After cooling, this produces a vitreous phase layer with sintered molybdenum powder strongly adhering to the alumina surface. A nickel layer is often added to this metallization to improve the brazing with conventional filler alloys.

Brazing between alumina and a metal induces a stress at the level of the join due to the thermal expansion mismatch. Kovar or Dilver are the commercial names of Fe-Ni-Co alloys developed for brazing on alumina. Their thermal expansion is close to alumina at brazing temperature. Depending on their thermal expansion and also their yield strength, metals and alloys are more or less suitable for brazing on alumina. The best is niobium, having a similar thermal expansion and low yield strength. Copper is also acceptable, even if the thermal expansion is high, the yield strength is low. Stainless steel having both high thermal expansion and high yield strength should be avoided.

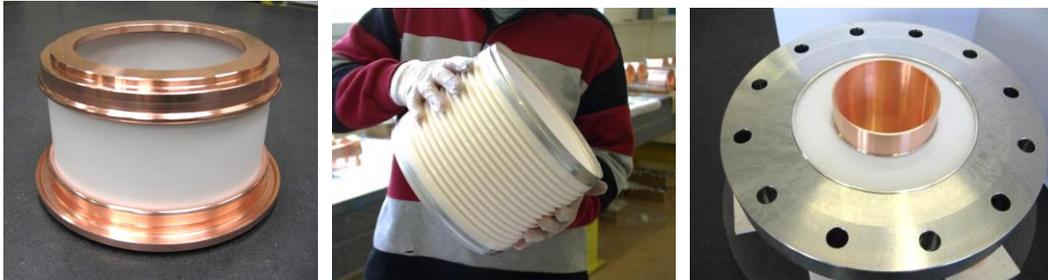

**Fig. 7:** Example of vacuum brazed alumina pieces with moly-manganese metallization: copper (left) and Kovar (centre) collar, titanium-alumina-copper assembly (right).

### 3.4.3 *Vacuum brazing of other ceramics*

The moly-manganese process is possible only for alumina. For other ceramics, we can use another technique called active brazing.

In this process, we use a metal filler alloy containing an 'active' metal. This metal is generally titanium, although it can be also be zirconium or beryllium, and is able to interact with the ceramic surface. At brazing temperature (above 800°C) and in 'good' vacuum (~ $5 \times 10^{-6}$ Torr), an active brazing alloy can form complex chemical components at the metal/ceramic interface. For example, a CuAgTi alloy on a SiC ceramic will form $Ti_3SiC_2$, $Ti_5Si_3C$, $Ti_5Si_3C_x$, etc. These intermetallics are brittle and the brazing parameters as well as the active element concentration must be carefully controlled. However, this method allows for the assembly of almost any ceramic on a metal.

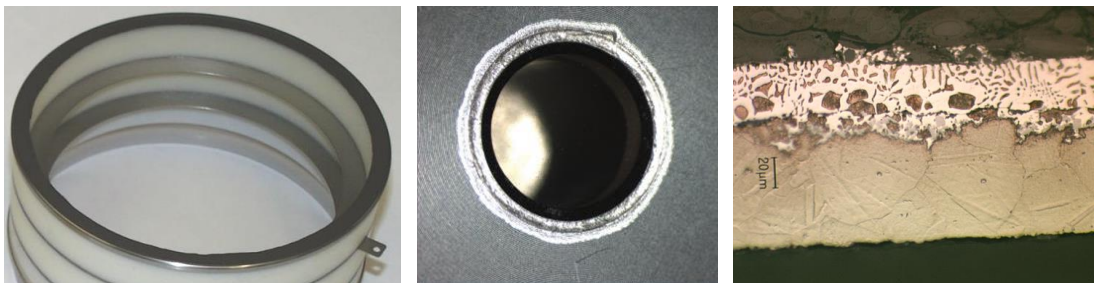

**Fig. 8:** Example of active brazing: sapphire-Nb-Alumina (left), diamond-Ti (centre), carbon-CuNi (right)

### 3.4.4 *Vacuum soldering*

Vacuum soldering is not commonly used but can be useful for the assembly of delicate components for non-baked-out vacuum systems. The solders can be Sn-Ag or Sn-Pb alloys with soldering temperatures around 240°C and 200°C respectively. High purity solders without high vapour pressure elements are in any case mandatory.

The wetting of the liquid metal is much less efficient that in brazing. It can be sufficient however on a Cu or Ag layer.

## 4 Present trends and future perspectives

There are many trends and future perspectives in the manufacturing of components for vacuum and accelerator applications. One can mention the CAM (computer-aided manufacturing) which allows for complex control of the machine centre for unconventional shape machining. For example, vacuum flanges with a knife and a rectangular or more complicated profile.

Metal additive manufacturing allows for unlimited shapes to be manufactured. The variety of materials available is also continuously expanding, and includes aluminium, stainless steel, and titanium alloys. The presence of impurities and porosity is an issue for vacuum applications however and this process is still under qualification.

High-velocity forming processes have been brought up to date for the manufacture of accelerating cavities such as electrohydraulic forming and explosive forming. Geometrical precision, better surface finishing, and reproducibility are the main advantages of these techniques.

# 5  Concluding remarks

High vacuum component manufacturing needs high quality materials. This is especially important when heat treatments, including vacuum firing, welding, or brazing, are required. The majority of leaks observed at welds are due to problems with the base materials.

The manufacturing process is defined by the design, however, the design must be defined first by the optimum available manufacturing process.

Trapped volumes in welding and the effect of the heat treatment in vacuum brazing are example of issues to keep in mind when designing the assembly. On the other hand, some established restrictive rules such as non-filler metal for TIG or no brazing between vacuum and water deserve to be re-examined.